\documentclass[twocolumn,american,aps]{revtex4}
\usepackage{pslatex}

\begin{document}

\title{Exclusion of correlation in the theorem of Bell}

\author{A. F. Kracklauer }

\thanks{kracklau@fossi.uni-weimar.de }

\begin{abstract}
Several fatal defects in recent defenses of Bell's theorem are identified. It
is shown again that ``proofs'' of the existence of non-locality are not valid
because they inadvertently exclude all correlation. A fully classical simulation
of EPR-B correlations, based on using Malus' Law for modeling both photocurrent
generation and the ``coincidence circuitry,'' is described. 
\end{abstract}
\maketitle
Bell asserted that:

\begin{quote}
... a hidden variable interpretation {[}of Quantum Mechanics (QM){]} has indeed
a grossly non-local structure ... characteristic, according to the result {[}he
``proves''{]}, of any such theory which reproduces exactly the quantum mechanical
predictions.\cite{JB}
\end{quote}
And again:

\begin{quote}
In a theory in which parameters are added to quantum mechanics to determine
the results of individual measurements, without changing the statistical predictions,
there must be a mechanism whereby the settings of one measuring device can influence
the reading of another instrument, however, remote. Moreover, the signal involved
must propagate instantaneously, so that such a theory could not be Lorentz invariant.\cite{JBb}
\end{quote}
These are categorical assertions and informally have been denoted, understandably,
as a ``theorem.'' What is not immediately evident in the above quotations,
is that these statements were made in connection with a very particular type
of experiment, namely those suggested by Einstein, Podolsky and Rosen (EPR)\cite{EPR}
but modified by Bohm (EPR-B)\cite{DB}. Bell formulated his arguments, ostensibly
facilitating empirical proof of these assertions, explicitely in terms of this
type of experiment.\cite{JB}

Recently in Ref. \cite{HP}, Bell's arguments have come under renewed attack
by Hess and Philipp (HP) who observe that in fact Bell in his ``proof'' neglected
to consider the possibility of time-dependant correlations in EPR-B experiments.
The development of their criticism is not simple and, moreover, they have not
given a transparent example or model of an actual EPR-B experiment involving
such time-dependant correlations.

In response to Ref. \cite{HP} and in defense of Bell's arguments, Gill, Weihs,
Zeilinger and \.{Z}ukowski (GWZZ)\cite{GWZZ} have brought HP's criticism itself
under attack; and, they further claim to have found a reformulation of Bell's
``proof'' that brings out ``very precisely the assumptions behind the theorem
of Bell.'' Likewise, in reaction the Ref. \cite{HP}, Mermin has defended his
rendition of Bell's argument. He claims to have: 

\begin{quotation}
...put forth {[}a{]} special case of Bell's theorem over twenty years ago to
demonstrate to non-scientists in a simple but rigorous way precisely what was
extraordinary about quantum correlations. ...its transparency also makes it
a good testing ground for claims of conceptual error in the formulations or
proof of Bell's theorem. Confusion buried deep in the formalism of very general
critiques tends to rise to the surface and reveal itself when such critiques
are reduced to the language of my very simple example.\cite{NDM}
\end{quotation}
In view, however, of the existence of counterexamples to Bell's theorem in the
form of local realistic models of exactly those experiments testing Bell's assertions
quoted above\cite{AB}\cite{AKa}, which both GWZZ and Mermin ignore, their
claims must be fundamentally erroneous. Herein some aspects of these authors'
arguments are dissected to see exactly where they fail. Although counterexamples,
as a matter of formal logic, settle the issue of the validity of Bell's so-called
theorem, many are unconvinced by global criticisms, and covet detailed analysis
of just where errors occurred. Some of that detailed analysis follows below.

\section{GWZZ's ``proof'' of Bell's theorem}

GWZZ's approach is not the standard EPR-B formulation. The first and most obvious
difference is that GWZZ consider a simulation that involves only dichotomic
variables as inputs and outputs; i.e., functions that take on only the values
of \( \pm 1 \), in contrast to the continuous polarization settings over the
\( [0,\; 2\pi ] \) range of polarizer settings considered usually in EPR-B
experiments. Further, GWZZ do not stipulate precisely how what they label the
``photons'' emitted in opposed pairs are correlated. In so far as the correlations
of the pairs in EPR-B experiments are the very source of the phenomena that
Bell considered should serve to expose the preternatural character of QM, such
carelessness in what GWZZ call ``our own proof of Bell's theorem'' is more
than just negligent; it admits, as shall be shown, an egregious error. Additionally,
they specify that, what they label the ``polarizers,'' are to be set independently
and randomly to only two settings based on coin tosses or other genuinely random
procedures. These differences with customary proofs of Bell's theorem all introduce
additional confusions and errors that exacerbate those already present in Bell's
formulation.

Finally, GWZZ's encoding of ``locality'' is not new, by their own acknowledgment,
and therefore, as suspect as the encoding used by Bell himself, which turned
out to be misdirected, as was demonstrated conclusively by the construction
of counterexamples. In short, we hold that the GWZZ ``proof'' is riddled with
confusions and errors, both the old and new.

\section{Binary variables}

A pervasive confusion throughout Bell's analysis is that concerning the role
of individual outcomes, versus the density of outcomes per unit time as a function
of angle\cite{AKb}. Bell's discussion of that EPR-B experiment he envisages,
leads the reader to think that he is considering the correlation of the individual
coincident events. In fact, however, all the expressions he writes do not pertain
to individual events, but to the relative frequency of coincidences at given
settings of the detectors per unit time. This turns out not to be just an accident,
but a necessity, as QM in fact does not predict the precise values of individual
events (nor correlations derived from such), but just their expectations given
a particular circumstance or experimental setup. That is, while QM provides,
at best, the spectrum and relative frequencies of the observable quantities,
it does not predict any particular outcome. Thus, for an EPR-B experiment, QM
only enables one to calculate the average, or ``expectation value'' of the
number of coincidences per unit time given the polarizer settings. Moreover,
this is exactly what is measured in the laboratory in the form of a photo\-current
intensity comprising what is in effect the average of many individual photoelectrons
which is then plotted as a function of the difference in the polarizer settings. 

If, after whatever simulations GWZZ carry out, they seek to compare the result
with those averages or ``expectations'' that can be obtained by calculation
from QM, as demanded by the logic of Bell theorems, then certainly their simulation
must produce entities that have the same physical units as those resulting from
the QM calculations. The GWZZ ``proof'' fails this requirement; it deals exclusively
with correlations of individual events. 

It can be shown, moreover, that any four dichotomic sequences, with zero mean
and values \( \pm 1 \), no matter how derived or how correlated, tautologically
satisfy a CHSH type Bell inequality\cite{AKc}, so the correlations considered
by GWZZ can not distinguish between alternatives, because there can be none
at this level. This, however, is not the main point herein, which is only that
GWZZ's ``proof'' of Bell's theorem does not yield the sort of items that can
be compared with a correspondent from QM.

\section{Input, output variables and labels}

The coincidence probabilities involved in analysis of EPR-B experiments take
the form \( \rho (a,\, b,\, \lambda ) \) where \( a \) and \( b \) specify
the polarizer settings, \( \lambda  \) is a ``hidden variable'' that presumably
specifies the state of the pairs (or twins) and \( \rho  \) is a density or
ratio of the number of times a coincidence event is registered when its arguments
have particular values over the total number of coincidences for all values
of the arguments. Just what all this means must be parsed and specified with
great care. The arguments, or independent variables in \( \rho  \), for example,
strictly speaking should have the units of angular displacement, radians or
degrees, in so far as in the standard experiments they correspond to polarizer
settings. However, in some particularly simple cases, when only a few angles
are considered, then the actual values; e.g., \( 0 \) or \( \pi  \) radians,
can be given labels which are meaningful in the context of the experiment, \emph{up}
and \emph{down}, say. This can severely confuse the mathematics as \emph{up}
and \emph{down} do not lend themselves to symbolic manipulation according to
the usual rules of algebra, etc. This is sometimes resolved by using numerical
labels such as \( -1 \) and \( +1 \), for example, which do lend themselves
to mathematical manipulation---however, without automatically vesting the results
with meaning! One can not just throw out physical units or introduce numerical
labels for the convenience of arithmetic. These same considerations apply in
spades to the significance or values of \( \rho (a,\, b,\, \lambda ) \), which
are never just \( \pm 1 \)'s, (labels), but ratios of coincidence counts or
densities of counts per unit time; the physical parallel of probabilities. Thus,
GWZZ's proposal to compute the correlation of the \emph{labels} of the outcomes,
without specifying what this should mean (\emph{inter alia} by giving the physical
units), introduces pervasive confusion.

Further, Bell and many who followed, never in their writings distinguished between
the active and passive input variables. Bell argued, for example, that locality
requires that the output on one side can not depend on the polarizer setting
on the other side. The fact is that the ``hidden parameter(s)'' \( \lambda  \)
is(are) to specify the state of the twins. Thus, when the pair has a suitable
state (correlation, or value of \( \lambda  \)), it transfers this property,
so to speak, through the polarizers, which, if set appropriately given the value
of \( \lambda  \), will result in \emph{coincident} output events at the detectors.
Obviously, the output on one side (perhaps labeled by the polarizer setting
at which it occurred) is independent of the \emph{passive setting} on the other
side; but, the occurrence of \emph{coincidence} detections on both sides, in
conjunction with the \emph{active} inputs from the twins does depend on the
passive settings on both sides. It is the twins that are correlated, and do
the ``transfer,'' not the passive polarizers. In addition, throughout the
analysis, the ``sample space'' consists of \emph{coincidence events,} not
individual events which are deliberately neglected in the calculations and explicitly
excluded by coincidence circuits in the experiments. The confusion on this point
in GWZZ's formulation is vast; it is not specified how the photons must be correlated,
nor what is to be measured as a coincidence, so that a relationship to EPR's
and Bell's reasoning is moot.

\section{QM non-locality}

The definition of ``locality'' used for the GWZZ argument does not encompass
the realities of QM as ensconced in super-position and wave function collapse
or the ``projection hypothesis.'' In his analysis Bell also did not explicitly
delineate the possible source of non-locality in QM. Most readers infer that
he in fact understood this feature and just glossed over it out of familiarity.
But this is a serious defect in his presentation which ultimately led him into
error.

All correlation of material objects which are space-like separated are in a
sense ``non-locally'' interrelated. This is, however, not in conflict with
Einstein's stipulation that no influence can transpire faster that the speed
of light in vacuo, if the objects obtained their correlation at some common
event in both of their past light cones. This is the point at which QM introduces
possible non-locality. QM holds that each of these twins is ontically ambiguous
and comprised of a ``superposition'' of exclusive outcomes until the moment
of measurement when a particular value, i.e., just one of the outcomes incorporated
in the superposition, is ``projected'' out. This is the oft remarked ``collapse''
of the wave function. Now, it is here that non-locality arises when a measurement
of one of the correlated twins is made, because by symmetry it must be that
the other twin's wave function also collapses, and this must happen at precisely
the same instant as the measurement on the first particle.

In essence, what Bell claims to have demonstrated, is that it is impossible
to account for EPR-B correlations without reference to such wave collapse. \emph{Counterexamples
conclusively demonstrate that this is not so!} These correlations can easily
be taken into account in fully local (thence realistic) models\cite{AB}\cite{AKa}.
(For a local-realistic model of an EPR-B experiment with time dependent correlations
as considered by HP, see \cite{11}.)

To encode locality in the relevant formulas, Bell supposed that the QM correlation
of EPR-B measurements, \( P(a,\, b) \), in terms of the individual, event-detection
probabilities of a deeper theory, should be rendered as follows: 
\begin{equation}
\label{bellcor}
P(a,\, b)=\int A(a,\, \lambda )B(b,\, \lambda )\rho (\lambda )d\lambda ,
\end{equation}
 where \( \rho (\lambda ) \) represents some possible density over the variable
set \( \lambda  \) from the deeper, `hidden,' theory. Bell described \( A(a,\, \lambda ) \)
and \( B(b,\, \lambda ) \) as ``the result of measuring \( \sigma _{1}\cdot a \)
and \( \sigma _{2}\cdot b \)'' and \( P(a,\, b) \) as the ``expectation
of the value of the product of \( \sigma _{1}\cdot a \) and \( \sigma _{2}\cdot b \),''
i.e., their correlation. If these definitions are to be mutually consistent
among themselves and with their QM correspondents, then we may write (after,
if need be, exchanging labels for values): \( A(a,\, \lambda )=a\rho _{A}(a,\, \lambda ) \)
and \( B(b,\, \lambda )=b\rho _{B}(b,\, \lambda ) \) so that Bell's joint probability
would therefore necessarily have to be of the form: 
\begin{equation}
\label{belljoint}
\rho (a,\, b,\, \lambda )=\rho _{A}(a,\, \lambda )\rho _{B}(b,\, \lambda )\rho (\lambda ),
\end{equation}
 where all \( \rho  \)'s are probability densities over the appropriate parameter
spaces.

The crucial feature incorporated in Eq.~(\ref{belljoint}) is Bell's encoding
of locality, namely, that the probability of a photon detection at station \( A, \)
\( \rho _{A}(a,\, \lambda ), \) must be independent of the settings of the
measuring apparatus at station \( B \), i.e., that it should not depend, he
says, on the variable \( b, \) and visa versa. Using Eq.~(\ref{bellcor}) then,
Bell and others derived inequalities for sets of correlations, \( P \)'s, which
are intended to be empirically testable; e.g.: 
\begin{equation}
\label{BI}
|P(a,\, b)-P(a,\, b')|+|P(a',\, b')+P(a',\, b)|\leq 2.
\end{equation}

According to basic probability theory, however, joint probabilities, when expressed
in terms of the probabilities of \emph{individual} detections at stations \( A \)
and \( B \), are encoded according to Bayes' formula, a.k.a. the ``chain rule'':
\begin{equation}
\label{feller}
\rho (a,\, b,\, \lambda )\equiv \rho _{A}(a|\, b,\, \lambda )\rho _{B}(b|\, \lambda )\rho (\lambda ),
\end{equation}
 where \( \rho (c|\, d) \) denotes a \emph{conditional} probability; i.e.,
the probability of the occurrence of an event parameterized by \( c \) given
that a \emph{condition} specified by \( d \) is met. In application to the
EPR-B experiment, for example, \( \rho _{A}(a|\, b, \)\( \,  \)\( \lambda  \))
is the probability of a detection of a photon at station \( A \) when its polarizer
is in the \( a \) direction, given that its companion photon is detected coincidently
at station \( B \) when its polarizer is in the \( b \) direction. The form
of this equation is somewhat arbitrary; the interdependencies of \( \rho _{A} \)
and \( \rho _{B} \), in other words the correlations, can, with no untoward
consequence, be built into either \( \rho _{A} \) or \( \rho _{B}. \) In any
case, there is no implication of non-locality in such a condition; the correlation
is imbued by a ``common cause'' in the past light cones of both entities,
and carried, so to speak, by the active input \( \lambda  \), and not by the
passive polarizer settings.

With respect to Bell's analysis, the critical point here is: the right side
of Eq.~(\ref{feller}) reduces to the integrand of Eq.~ (\ref{bellcor}), Bell's
basic supposition, if and only if: 
\begin{equation}
\label{iff}
\rho _{A}(a|\, b,\, \lambda )\equiv \rho _{A}(a|\, \lambda )\; \; \forall b.
\end{equation}
 If this is to encode locality, then one must presume that the presence of \( b \)
in the conditional probability \( \rho _{A}(a|\, b,\, \lambda ) \) implies
that this correlation necessarily involves superposition and projection. But
this is not necessary. Ordinary correlation invested in the twin objects at
any point in time in the past light cones of both twins, e.g., at their birth,
suffices. This relationship is then exposed by appropriately set detectors,
whose `settings' \( a \) and \( b \) play a purely passive role. They can
be set at any time early enough to be in place when the twins arrive. This is
exactly the tactic that permits the construction of local models of all EPR-B
experiments. Here it can be seen clearly that Bell's encoding inadvertently
precludes \emph{all} correlation, contrary to EPR-B's hypothesis. By implicitly
endorsing Eq. (\ref{iff}), GWZZ, following Bell, simply repeat his error. 

It can be shown, moreover, that by correctly taking correlation into account,
Bell-inequalities take the form:

\begin{equation}
\label{BItrue}
|P(a,\, b)-P(a,\, b')|\leq 2,
\end{equation}
which is a scarcely surprising inviolable tautology\cite{AKc}. This, however,
is not the main point herein, which is only that GWZZ's ``proof'' of Bell's
theorem is based on invalid assumptions.

This defect in Bell's argumentation was spotted, apparently, for the first time
by Jaynes as early as 1988\cite{EJ}. Subsequently it has been rediscovered
independently by at least five others\cite{EJ}-\cite{BL}, although ignored
by GWZZ and like-minded non-locality proponents. It is this writer's contention
that it is actually the core of HP's argument too. In this context, it should
be observed that HP's argument can be weakened, as the counterexamples demonstrate,
in the sense that all that is needed to demolish Bell's logic is the observation
that the correlations are invested in the past light cones of both detection
events, not that, as HP seem to hold, that the correlations must be time variable.

\section{Loopholes}

Separate from the soundness, or lack thereof, of Bell's analysis, critical arguments
concerning experiments done to test Bell's hypothesis have been advanced to
the effect that technical factors provide ``loopholes'' to evade Bell's conclusions.
One of the most discussed is that known as the ``detection'' loophole, which
arises in experiments, because real detectors are not perfectly efficient and
may not register all pairs emitted by the source. If this is the case, it is
possible, however unlikely, that the detected set of coincidences might not
be a fair representation of the whole set, but rather be skewed such that the
statistics of the measured set gives results at odds with the statistics of
the whole set. In effect, this problem arises because, experimentally one can
not determine exactly the denominator of the various ratios that constitute
the probabilities of interest.

Another such loophole can arise in principle if the twins can communicate with
each other and, in effect, carry out a ``conspiracy.'' In this case, the twins,
perhaps via as yet some totally unknown effect, might skew the statistics again
to support false conclusions. Analysis of these loopholes has lead to the suggestion
that the latter, ``conspiracy,'' loophole, and perhaps the former too, can
be corked by randomly setting the polarizers in an EPR-B experiment at a time
so close to the detection that it would be impossible, on account of detector
separation, for the twins to collaborate using light signals. This is the rationale
behind using randomly set polarizers, as GWZZ propose, unstated though it is
in Ref. \cite{GWZZ}.

In no case, however, does the possible existence of a loophole in the experiments
address the fundamental correctness of Bell's ``theorem'' per se, which concerns
the statistics of the ideal case with uncorrupted outputs. It is at this point
that GWZZ's formulation introduces another fundamental confusion. In the first
place, unless the ``conspiracy'' mechanism itself, or another bias, is encoded
into a simulation, logically there is no need to program countermeasures. Their
new ``proof,'' using randomly set polarizers, is therefore unnecessarily complex,
because the point made by HP addresses the validity of the theorem, not the
logical tightness of the experiments against loopholes. The issue is not, as
GWZZ assert, whether the experimenters have ``freedom'' to set the polarizers,
so as to prevent conspiracies, but rather, whether EPR-B correlations from idealized
experiments can be modeled without recourse to non-local interaction, contrary
to Bell's claims.

\section{Counterexamples}

If Bell's theorem is wrong, and if there is no QM in the spaces in which the
EPR-B experiments have been formulated, as argued above, then certainly it should
be possible to model those experiments using only concepts from classical physics.
This is indeed so; consider the following:

The model described below consists of simply rendering the source mathematically,
and a computation of the coincidence rate. Photodetectors are assumed to convert
continuous radiation into an electron current at random times with a Poisson
distribution, but in proportion to the intensity of the radiation. The coincidence
count rate is taken to be proportional to the second order coherence function. 

The source is assumed to emit a double signal for which individual signal components
are anticorrelated and confined to the vertical and horizontal polarization
modes; i.e.
\begin{equation}
\label{source}
\begin{array}{cc}
S_{1} & =(cos(n\frac{\pi }{2}),\: sin(n\frac{\pi }{2}))\\
S_{2} & =(sin(n\frac{\pi }{2}),\: -cos(n\frac{\pi }{2}))
\end{array}
\end{equation}
 where \( n \) takes on the values \( 0 \) and \( 1 \) with an even random
distribution. The transition matrix, \( \chi  \), for a polarizer is given
by,

\begin{equation}
\label{polarizer}
\chi (\theta )=\left[ \begin{array}{cc}
\cos ^{2}(\theta ) & \cos (\theta )\sin (\theta )\\
\sin (\theta )\cos (\theta ) & \sin ^{2}(\theta )
\end{array}\right] ,
\end{equation}
 so the fields entering the photodetectors are given by:
\begin{equation}
\label{output}
\begin{array}{cc}
E_{1} & =\chi (\theta _{1})S_{1}\\
E_{2} & =\chi (\theta _{2})S_{2}
\end{array}.
\end{equation}
 Coincidence detections among \( N \) photodetectors, \( \gamma  \), (here
\( N=2 \)) are proportional to the single time, multiple-location, second-order
cross correlation, i.e.:
\begin{equation}
\label{e2}
\gamma (r_{1},\, r_{2},..r_{N})=\frac{<\prod ^{N}_{n=1}E^{*}(r_{n},\! t)\prod ^{1}_{n=N}E(r_{n},\! t)>}{\prod ^{N}_{n=1}<E^{*}_{n}E_{n}>}.
\end{equation}
 It is easy to see that for this model the denominator usually consists of factors
of \( 1 \). The final result of the above is:
\begin{equation}
\label{EPRcor}
\rho (\theta _{1},\theta _{2})=\frac{1}{2}\sin ^{2}(\theta _{1}-\theta _{2}).
\end{equation}
 This is immediately recognized as the so-called `quantum' answer. (Of course,
it is also Malus' Law.) Eq. (\ref{EPRcor}) is the result for like channels.
A similar expression with the sine replace by cosine, pertains to unlike channels.
The total correlation is then \( \{P(+,\, +)+P(-,\, -)-P(+,\, -)-P(-,\, +)\}/\{P(+,\, +)+P(-,\, -)+P(+,\, -)+P(-,\, +) \)\}for
which the result here is \( -cos(2(\theta _{1}-\theta _{2})) \), as is found
also using QM. 

Note that the mere existence of such models constitutes, irrespective of other
critiques, counterexamples undermining the validity of Bell's theorem. An informal
criticism of this model is that actually it incorporates quantum structure in
a covert manner. This is fully refutable on the basis that the model is not
formulated in either phase space or quadrature space (i.e., in terms of amplitude
and phase) which are the only two spaces in which the generators of translations
parallel to the axes are subject to Heisenberg Uncertainty between themselves.
(Noncommutivity among Stokes operators is not due to Heisenberg Uncertainty,
but to the structure of the three dimensional rotation group. This structure
enters only when the wave vector common to both polarization modes rotates;
this is an entirely separate, and fully geometric matter.)

The above model is unfortunately also a conceptual ``black box.'' That is, while
the inputs and outputs are clear, just what happens in detail inside the model
is obscure, not physically motivated. 

Early attempts to model EPR-B correlations in detail considered that the pairs
in each superposition emitted by the source elicit independent photocurrent
processes. For this pair the correlation, assuming that on each side the signal
falls on a photodetector and evokes photoelectrons proportional to its square;
i.e., its energy, would then be given by
\begin{eqnarray}
\frac{\frac{1}{2\pi }\int^{2\pi }_{0} \cos ^{2}(\theta -\lambda )\sin
^{2}(\theta -\rho )d\theta }{\sqrt{\frac{1}{2\pi }\int^{2\pi }_{0}\cos
^{2}(\theta -\lambda )d\theta \,\,\frac{1}{2\pi}\int^{2\pi }_{0} \cos
^{2}(\theta -\rho )d\theta }}= &  & \nonumber \\
\frac{1}{2}-\frac{1}{4}\cos (2(\lambda -\rho )). &  & \label{furry} 
\end{eqnarray}
 This is the explanation for EPR-B correlations proposed originally by Furry.
In his conceptions, he envisioned that as the two signals obtained some distance
from the source, they became independent or disentangled, thereby obviating
the role of nonlocal interaction. To some degree this same notion underpinned
the attempt by Jaynes and collaborators to advance the claim that a semiclassical
treatment of electromagnetic phenomena might completely or in large measure
preempt quantum electrodynamics. Early experiments by Clauser and collaborators
torpedoed this hope, however. They observed that the visibility predicted by
Eq. (\ref{furry}) was limited to 50\%, but that both QED and experiments yield
100\%. 

This would be the end of this story were it not for technicalities overlooked
in the derivation of Eq. (\ref{furry}). They include that in the experiments,
the currents considered actually differ from those implied above. In the experiments,
namely, the full current is not taken into account, but unpaired, single photoelectrons
observed in either arm of the experiment are excluded deliberately by coincidence
circuitry. Further, the geometric relationship used to construct the numerator
is incorrect---as will be shown below. Finally, the signals are not distributed
over \( 2\pi  \), but limited to the vertical and horizontal.

It is worthwhile, therefore, to simulate in detail exactly what does happen
in an EPR-B experiment. This can be done as follows: The source is assumed to
emit paired, oriented (i.e., either vertically or horizontally polarized) pulses
that are anticorrelated, e.g., left::vertical plus right::horizontal and visa
versa. Then each pulse in each wing is directed through a polarizer set at an
angle \( \theta _{l,\, r} \) after which the emerging signal is sent through
a polarizing beam splitter (PBS) whose axis is parallel to that of the source.
Each arm of the PBS is taken to be observed by a photodetector, of which there
are four, two on each wing. Each photodetector is independent of all others;
each generates its own photocurrent where the intensity is proportional to the
intensity of the total field impinging on it, in other words, according to Malus'
Law. Each photocurrent is comprised of photoelectrons for which the arrival
time is a Poisson random variable. Positive coincidences are those between like
channels (e.g., vertical:left---vertical:right), negative coincidences are between
unlike channels. 

Results from the simulation show that: 1) correlations among the photoelectrons
considered as Poisson processes stimulated according to Malus' Law always respect
Bell inequalities; 2) correlations between the current \emph{densities,} not
the Poisson processes in the photodetectors, defined between channels using
Malus' Law again, conform with those calculated using QM, and as is well known,
do not respect Bell inequalities. 

The latter coincidences are exactly those selected by ``coincidence circuitry,''
which, in effect, upon detecting a photoelectron in any one detector, then looks
for coincident photoelectrons in the other detectors within a specified window,
that is, with respect to itself, not the source. The test for such a coincidence
involves an expression that can not be factored, as indeed probabilities for
coincident events can not be factored into the product of probabilities of independent
events. In this case, however, this in no way implies nonlocality because the
test has nothing to do with the existence of a photoelectron or any other physical
interaction, just its relative position to other, already existing photoelectrons.
Photoelectrons that fail the test do not suffer any physical consequence, they
are just ignored when identifying coincidences---as ``coincidence circuitry''
ignores, but does not otherwise affect unpaired photoelectrons. In the end,
coincidence of \emph{intensities} between output channels can not be expressed
as the direct product of the intensities generating the photocurrents for the
same reason \( \cos (\theta -\phi )\neq \cos (\theta )\cos (\phi ) \), rather
\( \cos (\theta )\cos (\phi )+\sin (\theta )\sin (\phi ) \), a simple trigonometric
truth. The crucial insight from the simulation is that the role of coincidence
filtering is at the core of EPR-B correlations; it is really the effect being
studied, which is, at root, just geometry. That is, the simulation makes it
clear that the statistics of the Poisson processes in the photodetectors are
not the substance of the coincidence statistics measured by the coincidence
circuitry, rather just the geometric relationship, i.e., Malus' Law, between
the \emph{intensities} in the output channels.

This model and simulation are directly extendable to all other tests of EPR-B
correlations, including multi-particle ``GHZ'' experiments. See \cite{AKa}.
To obtain Mermin's ``special case,'' which is in fact just the ordinary case
restricted to considering only three angles, simply set \( \theta _{1}-\theta _{2}=0,\; 2\pi /3,\; 4\pi /3 \).
His stipulations i.) and ii.) in ref. \cite{NDM} are encoded in Eqs.~(\ref{source})
and (\ref{e2} ) respectively. The extreme perplexity he finds in this case
appears to derive from an implicit attempt to attribute the random input of
the environment at the photodetectors, encoded in Eq. (\ref{e2}) as Malus'
Law and giving a continuous output, to the random aspect of the dichotomic source,
Eq. (\ref{source}). Furthermore, he fails to take account altogether of the
coincidence circuitry and its role in selecting the sequences that exhibit perplexing
statistics.

Likewise, GWZZ et al.'s ``proof'' is also just the standard EPR-B setup restricted
to two observation angles. In their explication of their proof, they also explicitly
go to considerable length to establish that the settings of the measurement
devices, \( A \) and \( B \) in their notation, are statistically independent
of the outcomes, \( X_{1},\: X_{2},\: Y_{1},\: Y_{2} \). This requirement parallels
exactly Bell's misconstrual in so far as he inadvertently encodes ``locality''
as the stipulation that the measurement settings are to be independent of the
outcomes. On this point, it might reasonably be questioned, however, what purpose
a measurement should have when the settings are independent of the outcomes;
why make such a measurement even?

\section{Conclusion}

Extraordinary claims deserve extraordinary proof. Non-locality, interchanging
the order of cause and effect as it can, is extraordinary in an extreme sense!
To begin, there is \emph{no} empirical verification in any area of science.
What is considered proof, is simply the coincidence of the statistics of certain
QM and empirical correlations. In view of the fact that the conclusion depends
on the \emph{interpretation} of these statistics, it clearly deserves prudent
scepticism and extensive reanalysis before being taken as fact.

Moreover, counterexamples are the nuclear weapons of logic, as it were. The
smallest, technical or artificial counterexample utterly devastates even the
most elaborate theorem. The state of play, therefore, is now such that proponents
of Bell's analysis or the logical viability of non-locality, to support their
case, must show that extant counterexamples are wrong or irrelevant. So long
as the counterexamples stand, all other arguments are suspended.

GWZZ's views in Ref. \cite{GWZZ} and Mermin's in Ref. \cite{NDM} respect neither
of these universal precepts.

As to the specifics of the GWZZ attack on HP's criticisms, the situation is
no better. HP's essential argument is against the validity of Bell's basic theorem,
not the experiments which could be vulnerable by fault of loopholes. Most of
GWZZ's counter criticisms, on the other hand, address issues that pertain to
evading the conspiracy loophole, not the theorem itself. The experimenters'
``freedom'' can only effect the settings of the measuring devices, which in
turn can only affect the time order in which the data is taken, not the actual
patterns in the data (presuming that a conspiracy was not deliberately encoded
into their simulation). GWZZ's further comments regarding the alleged covert
investment by HP of non-locality, is seen easily as a variation of the confusion
of passive with active variables and their role in conditional probabilities
pertaining to \emph{coincident} events. Having failed to delineate how non-locality
would enter in the first place, they then again by default attribute it to any
and all correlation, whatever the source. For example, when GWZZ assert that
Eq. (16) in Ref. \cite{HP} implies non-locality, they indulge exactly this
confusion regarding the nature of correlation among coincident events. Thus,
either GWZZ's comments and arguments actually don't pertain to HP's arguments
in the first place, or they just reiterate Bell's original mistake. 

Furthermore, moving the Bell argument in general into the arena in which individual
events, instead of the density of events, introduces, fundamental obstacles.
It precludes the possibility altogether of comparing any calculated result with
that obtained using QM or obtained from experiments. This error is not indulged
exclusively by GWZZ, but, regrettably, also pervades much current writing on
Bell's theorem, EPR-B correlations and non-locality.

GWZZ's and Mermin's arguments, in short, once again repeat Bell's basic error
of presuming effectively that all EPR-B correlations are expressed via projection
from superposition states. Whereas Bell explicitly recognized the role of correlation
but inadvertently excluded it by misencoding locality, GWZZ's arguments fail
even to analyze its importance and possible source and thereafter to take it
appropriately into account. In conclusion, GWZZ and Mermin have neither demolished
HP's criticism of Bell's ``theorem,'' nor contributed to the elucidation of
the issues under debate on the validity of non-locality or its necessity in
an extention of QM involving ``hidden variables.'' They also completely overlook
both 1) the profound contribution of ``coincidence circuitry'' and a correct
geometric rendering of its effects to the observed phenomena, and 2) the distinguished
role of \emph{intensity} correlations distinct from their underlying Poisson
processes.

\section*{Acknowledgment}

I thank Barry Schwartz for extensive insightful remarks on Jaynes' criticisms
of Bell's reasoning and a critical reading of this manuscript.

\end{document}